\documentclass{ws-mpla}

\usepackage{graphicx}
\usepackage{amsmath}
\usepackage{amsfonts}
\usepackage{amssymb}
\usepackage{epsfig}
\usepackage{graphicx}
\usepackage{slashed}

\begin{document}

\markboth{Authors' Names}
{Instructions for Typing Manuscripts (Paper's Title)}

\catchline{}{}{}{}{}

\title{PHENOMENOLOGY OF LIGHT STERILE NEUTRINOS:\\
 A BRIEF REVIEW}
 

\author{ANTONIO PALAZZO}

\address{Max-Planck-Institut f\"ur Physik (Werner Heisenberg
Institut), F\"ohringer Ring 6, 80805 M\"unchen, Germany}



\maketitle

\pub{Received (Day Month Year)}{Revised (Day Month Year)}

\begin{abstract}
An increasing number of anomalous experimental results are emerging, which cannot be 
described within the standard 3-neutrino framework. We present a concise discussion 
of the most popular phenomenological interpretation of such findings, based on 
a hypothetical flavor conversion phenomenon of the ordinary ``active'' neutrinos into 
new light ``sterile'' species having mass $m \sim \mathcal{O}(1)$\,eV. 

\end{abstract}

\ccode{PACS Nos.: 14.60.Pq,14.60.St}

\section{Introduction}	

A long series of neutrino oscillation experiments performed in the last two decades has
established that neutrinos are massive and mix. In the standard 3-flavor framework the three flavor eigenstates ($\nu_e, \nu_\mu, \nu_\tau$) mix with three mass eigenstates
 ($\nu_1, \nu_2, \nu_3$) through a unitary matrix entailing three mixing angles
 ($\theta_{12}, \theta_{23}, \theta_{13}$) and one CP-violating phase $\delta$.  
All the three mixing angles are now known to be different from zero ($\theta_{12} \simeq 34^\circ$, 
 $\theta_{23}\simeq 39^\circ$ and  $\theta_{13}\simeq 9^\circ$), while the preferred value of 
the CP-violating phase lies around $\delta \simeq \pi$, although with low statistical significance\cite{Fogli:2012ua}.
In the 3-flavor scheme the neutrino oscillations are driven by two independent mass-squared differences,
$\Delta m^2_{12} = m_2^2 - m_1^2\simeq 7.5 \times10^{-5}\, \mathrm{eV}^2$ (known as the ``solar splitting'') 
and  $|\Delta m^2_{13}| = |m_3^2 - m_1^2| \simeq 2.4 \times10^{-3}\, \mathrm{eV}^2$ (dubbed as the ``atmospheric splitting"). 
The neutrino mass hierarchy  (i.e. the sign of  $\Delta m^2_{13}$) is currently unknown and,
together with the CP-violating phase $\delta$, is at the center of an intense campaign of new experimental
searches.

While the standard 3-flavor framework has been solidly established as the only one 
able to describe the huge amount of information coming from solar, atmospheric,
reactor and accelerator neutrino experiments, a few ``anomalous'' results have emerged
in very-short-baseline (VSBL) neutrino oscillation measurements and in cosmological data 
analyses, which cannot be accommodated in such a scheme. 
The most popular interpretation of such anomalies is based 
on a simple extension of the 3-flavor paradigm, involving new additional light neutrinos 
(with mass in the eV range) which mix with the ordinary neutrinos. From the LEP measurement of the invisible decay
width of the Z boson\cite{ALEPH:2005ab}, we know that there are only three light (with mass below one half of the Z boson mass)
neutrinos which couple to the Z boson. Therefore, the new putative
light neutral fermions must be ``sterile'', i.e. singlets of the Standard Model gauge group,
to be contrasted with the ordinary ``active'' neutrino species, which are members of weak isospin doublets.
 
From a theoretically standpoint, it seems natural to expect the existence of such new gauge singlets
as they appear in many extensions of the Standard Model.  Indeed, the most popular models
of neutrino mass-generation, the so-called see-saw mechanisms, normally involve sterile neutrinos.
Although the majority of such extensions entail sterile neutrinos with mass close to the grand unification 
scale or the TeV scale,  {\it a priori} there is no theoretical constraint on the mass of these particles.
In fact, several models have been investigated in which much lighter sterile neutrinos arise (see the overview
given in\cite{Abazajian:2012ys}).  In essence, the theory only tells us that  sterile neutrinos {\em can} exist, 
without giving any certain information on their number and their mass-mixing properties, which 
ultimately have to be determined by the experiments.

From a phenomenological perspective, the sterile neutrinos must be
introduced without spoiling the basic success of the standard 3-flavor framework. 
This can be achieved in the so-called $3+s$ schemes, where $s$ new 
mass eigenstates are assumed to exist, separated from the three standard
ones by large mass splittings, with a hierarchical  spectrum
$|\Delta m^2_{12}| \ll  |\Delta m^2_{13}| \ll |\Delta m^2_{1j}| (j = 4, ..., 3+s)$.
This ensures that the oscillations induced by the new mass-squared differences
are completely averaged in the experimental setups sensitive 
to $\Delta m^2_{12}$-driven (solar) and  $\Delta m^2_{13}$-driven (atmospheric)
transitions, leaving unaltered the well-established frequencies of the standard oscillation processes.
With the additional assumption that the admixtures among
the actives flavors and the new mass eigenstates ($\nu_4, ..., \nu_{3+s}$) 
are small ($|U_{ej}|^2, |U_{\mu j}|^2, |U_{\tau j}|^2 \ll 1, j = 4, ..., 3+s$) 
[or equivalently that the new neutrino mass eigenstates are mostly sterile 
($|U_{s4}|^2, ..., |U_{s, 3+s}|^2\simeq 1$)],  the $3+s$ schemes leave almost untouched 
also the standard oscillation amplitudes. 

In such a way, the $3+s$ frameworks realize a peculiar enlargement of the standard
 3-flavor scheme, as they induce strong effects in those setups where the 
 anomalies show up,  while having only a small (but different from zero) impact on
 the ``ordinary'' pieces of data commonly  used for the 3-flavor global fits. 
However, we underline that, at the present level of accuracy one must take into account both
the leading effects normally expected in VSBL experiments and the small ones expected 
in the ordinary setups.

In what follows we concisely describe the anomalous results and discuss their 
mutual  (in-)consistency. We also provide a concrete example of how some ``ordinary'' 
pieces of data, namely the solar neutrino sector experiments together with the 
new dual-baseline  $\theta_{13}$-sensitive reactor experiments Daya Bay and RENO, 
are able to put interesting constraints on the 3+1 scheme. Finally we draw our conclusions.

\section{The anomalies}

\subsection{The accelerator anomaly}

Accelerator experiments with baselines $L$ of few tens of meters and neutrino energies $E_\nu$
of a few tens of MeV ($L/E_{\nu} \sim 1 ~{\mathrm m}/{\mathrm  {MeV}}$)  are sensitive probes 
of neutrino oscillations potentially occurring at $\Delta m^2 \sim 1\,\mathrm{eV}^2$.  Their results 
are commonly interpreted in terms of a new mass-squared difference $\Delta m^2$ and of an effective
mixing angle $\theta$. In a 3+1 framework the following identifications hold:  $\Delta m^2 \equiv \Delta m^2_{14}$ 
and $\sin^2 2\theta \equiv 4  |U_{e4}|^2 |U_{\mu4}|^2$.

\begin{figure*}[t!]
\vspace*{-0.0cm}
\hspace*{2.5cm}
\includegraphics[width=8.0 cm]{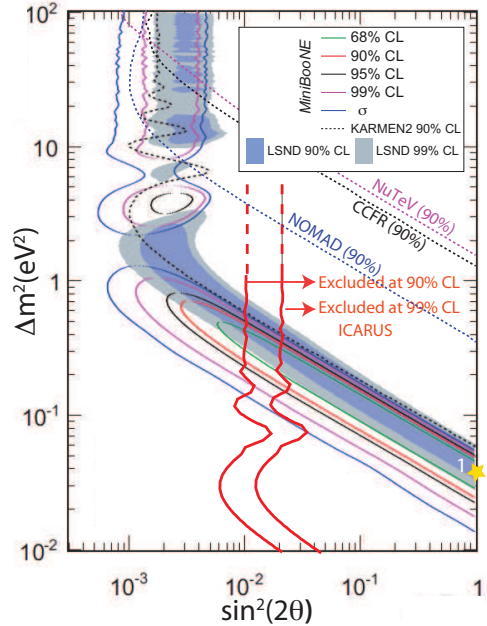}
\vspace*{0.25cm}
\caption{Regions allowed by the main published experiments sensitive to the accelerator 
anomaly superimposed to the limits established by the ICARUS experiment. Figure taken from\protect\cite{Antonello:2012pq}.
\label{fig1}} \end{figure*}  

In fact, the anomalous result  recorded at the LSND accelerator experiment\cite{Athanassopoulos:1996jb,Aguilar:2001ty}
was the first piece of data pointing towards light sterile neutrinos. Such an experiment, 
designed to study $\bar\nu_\mu\to \bar\nu_e$ transitions, evidenced an excess 
of electron antineutrino events a the $\sim3.8\sigma$ level\cite{Aguilar:2001ty}.
The mass-mixing regions preferred by LSND are depicted in Fig.~1 as colored bands. 

The experiment  KARMEN\cite{Armbruster:2002mp}, which is very similar to LSND, observed no such a signal,
but could not rule out all the mass-mixing parameter regions allowed by LSND, as shown
in Fig.~1, where the region excluded by KARMEN is delimited by the dashed gray line.
In particular, due to the smaller baseline (17.5 m vs 30 m) KARMEN could not exclude values of  
$\Delta m^2 < 2 ~{\mathrm {eV}}^2$ also leaving marginal room for a solution at
$\Delta m^2 \sim 7 ~{\mathrm {eV}}^2$ (see the combined analysis of LSND and KARMEN
performed in\cite{Church:2002tc}).

The experiment MiniBooNE, designed to test the LSND anomaly, and sensitive both to 
$\nu_\mu\to \nu_e$  and  $\bar\nu_\mu\to \bar\nu_e$ transitions, has given
differing results during the last few years. According to the latest data release\cite{AguilarArevalo:2012va} 
(differently from the past) MiniBooNE seems to lend support to the longstanding LSND anomaly. 
As a matter of fact, a combined analysis of $\nu_\mu\to \nu_e$  and  $\bar\nu_\mu\to \bar\nu_e$ channels
evidences an excess in the range $200 <E_\nu< 1250~$MeV at the $\sim3.8\sigma$ level.  
The mass-mixing regions compatible with this result are shown as colored contours in Fig.~1.   
It must be noted that most of the signal comes from the low-energy region $E_\nu < 475$~MeV
(not included in some of the previous analyses performed by the MiniBooNE collaboration),
where the background evaluation is particularly problematic.

An independent test of the LSND and MiniBooNE anomalies has been recently performed at the long-baseline
accelerator experiment ICARUS\cite{Antonello:2012pq}. This novel test has been possible because in
the ICARUS setup, due to the high energy of the beam ($<E_\nu> \sim 17$~GeV),  the 3-flavor effects induced by 
non-zero $\theta_{13}$ play a negligible role. As a result, the experiment is sensitive to sterile neutrino
oscillations, which due to the high value $L/E_\nu \sim  36.5$~m/MeV get completely
averaged, and appear as an energy independent enhancement of the expected rate of events.
However, ICARUS is not sensitive enough to rule out all the
relevant mass-mixing region, and could only restrict the allowed region to values of
$\sin^2 2\theta \lesssim10^{-2}$,   as shown in Fig.~1.   From the same plot  one can 
qualitatively infer that the combination of  the four experiments LSND, KARMEN, 
MiniBooNE and ICARUS, restricts the allowed mass-mixing parameters to
a small region centered around  ($\Delta m^2 \sim 0.5 ~{\mathrm {eV}}^2$, $\sin^2 2\theta \sim 5\times 10^{-3}$).

\subsection{The reactor and gallium anomalies}

The new refined calculations of the reactor antineutrino spectra recently 
performed in\cite{Mueller:2011nm,Huber:2011wv} have provided the main driving force 
for the renewed interest into light sterile neutrinos. These calculations
indicate fluxes which are $\sim 3.5\%$ higher than previous estimates\cite{Vogel:1980bk,VonFeilitzsch:1982jw,Schreckenbach:1985ep,Hahn:1989zr}
(corresponding to events  rates $\sim 6\%$ higher)  and have raised the so-called reactor
antineutrino anomaly\cite{Mention:2011rk}. In fact, adopting the new fluxes, the VSBL ($L\lesssim 100 $~m) reactor measurements 
show a clear deficit (a $\sim 3\sigma$ effect) with respect to the theoretical expectations.

Despite the  effort made by the authors of\cite{Mueller:2011nm}, 
who have included thousands of $\beta$-branches in the calculations, the new 
determinations are not entirely performed with an {\em {ab initio}} procedure.
In fact, approximately 10\% of all the $\beta$-branches remains 
unknown and their contribution is accounted for by adding 
up a few fictitious effective $\beta$-branches.  
The resulting overall (electron) $\beta$ spectrum is 
then ``anchored'' to that one measured by the ILL experiment\cite{VonFeilitzsch:1982jw,Schreckenbach:1985ep,Hahn:1989zr}
in the 1980s.
Therefore, at present, a systematic error in the ILL measurement
cannot be excluded as the origin of the reactor anomaly.

\begin{figure*}[t!]
\vspace*{-0.0cm}
\hspace*{1.10cm}
\includegraphics[width=10.0 cm]{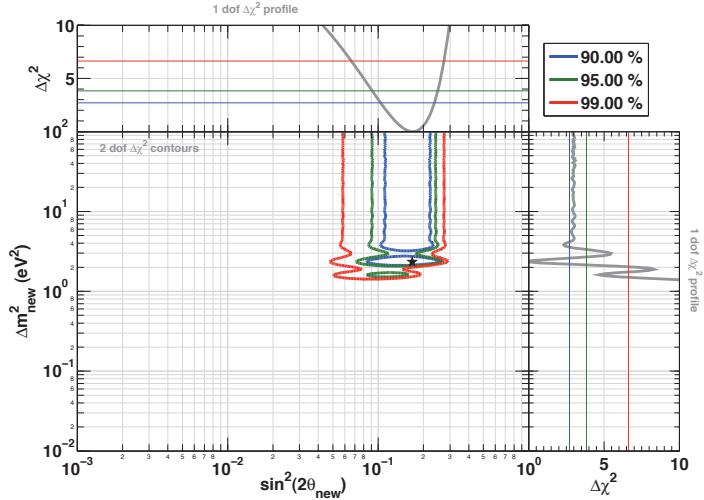} 
\vspace*{0.0cm}
\caption{Regions allowed by the combination of the reactor  and
gallium anomalies (figure taken from\protect\cite{Abazajian:2012ys}).\label{fig_re+ga}}
\end{figure*}  

An apparently unrelated deficit  has been evidenced in the calibration campaign conducted at the solar neutrino experiments
GALLEX and  SAGE\cite{Abdurashitov:2005tb,Giunti:2012tn} with high intensity radioactive sources. 
The exact statistical significance of the deficit fluctuates in the range $[2.7\sigma, 3.1\sigma]$ depending on
 the assumptions made on the theoretical estimate of the cross section $\nu_e + ^{71}\!{\mathrm {Ga}} \to ^{71}\!{\mathrm {Ge}} + e^-$ (see the discussion presented in\cite{Giunti:2012tn}). While the anomaly may well 
 represent a signal of new physics other (more prosaic) possibilities remain open such 
 as a systematic error in the ${\mathrm {Ge}}$ extraction efficiency or in the theoretical
 estimate of  the cross-section.
 
 Both the reactor and gallium anomalies can be interpreted in terms of a phenomenon of
 electron neutrino disappearance driven by sterile neutrino oscillations. 
 The result of a combined fit of the two anomalies 
 is shown in Fig.~2 in terms of a new mass-squared difference $\Delta m^2 _{new}$ 
 and of an effective mixing angle $\theta_{new}$. In a 3+1 framework the following 
 identifications hold: $\Delta m^2_{new} \equiv \Delta m^2_{14}$ and 
 $\sin^2 2\theta_{new} \equiv 4 |U_{e4}|^2 (1- |U_{e4}|^2)$. From Fig.~2 
 we see that values of $\Delta m^2_{new} \gtrsim 1\, \mathrm{eV}^2$ and relatively
 large values of $\sin^2 2\theta_{new} \sim 0.17$ are preferred (corresponding to 
values of $|U_{e4}|^2 \sim 0.04$).

\subsection{The dark radiation anomaly}

The latest cosmological data analyses of the Cosmic Microwave Background (CMB)
and large scale structure (LSS)\cite{Komatsu:2010fb,Hamann:2010bk,Giusarma:2011ex,Hamann:2011ge,Archidiacono:2011gq,Hamann:2011hu,Joudaki:2012uk,Wang:2012vh,Giusarma:2012ph}, show a weak ($\sim2 \sigma$) but consolidated 
trend towards extra relativistic degrees of freedom at the epoch of the CMB decoupling.
In Fig.~3 we report the results of the analysis performed in\cite{Hamann:2010bk}, 
where the number of extra relativistic species is denoted with  $N_s$.

Such hint for extra radiation,  often dubbed as ``dark radiation'', is not in contrast  with 
the constraints coming from the earlier epoch of big bang nucleosynthesis (BBN), 
which, however, allow for no more than one additional sterile species\cite{Izotov:2010ca,Mangano:2011ar}. 
It must be stressed that the bounds obtained in\cite{Izotov:2010ca,Mangano:2011ar}
are obtained under the hypothesis that sterile neutrinos were in thermal equilibrium 
with the primeval plasma prior to BBN (more precisely prior to the decoupling of the ordinary neutrinos).

The issue of thermalization of sterile neutrinos prior to the BBN epoch has been
recently reconsidered in\cite{Mirizzi:2012we,Hannestad:2012ky,Saviano:2013ktj}
(for earlier works see the list of references given in\cite{Abazajian:2012ys,Mirizzi:2012we}).
The basic result of these works is that for the mass-mixing parameters of current interest 
full thermalization of one sterile neutrino do occur unless a very large ($\gtrsim 10^{-2}$) 
initial lepton asymmetry is present. 

\begin{figure*}[t!]
\vspace*{0.2cm}
\hspace*{2.00cm}
\includegraphics[width=8.0 cm]{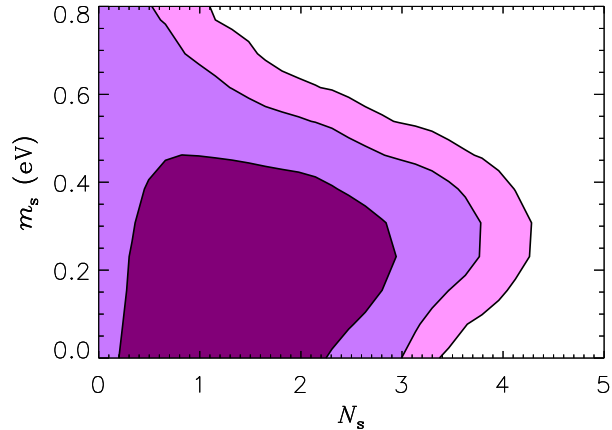}
\vspace*{0.2cm}
\caption{Regions allowed by the CMB data. The ordinary neutrinos are assumed
to be massless while the $N_s$ sterile species have a common mass $m_s$ (figure taken from\protect\cite{Hamann:2010bk}). 
\label{fig_cosmo}}
\end{figure*}  

New information on the radiation content of the early universe is expected to 
come during this year from the Planck mission. Awaiting  such results, one must
bear in mind that the excess of radiation is not specific of sterile neutrinos since it can be 
produced by alternative mechanisms (see for example\cite{Hasenkamp:2012ii} and references therein).
For this reason, a positive confirmation by Planck of the present hint would not necessarily imply the existence of sterile
neutrinos, although it would certainly reinforce such an hypothesis. On the other hand, a 
disconfirmation of the hint, while certainly disfavoring the hypothesis of sterile neutrinos, 
would not be able to definitely exclude this option, since some 
mechanism may suppress the contribution of the sterile neutrinos to the radiation content of
the early universe. Therefore, while the Planck results will give an important indication, it would be
prudent to leave the final word on sterile neutrinos to dedicated laboratory oscillation experiments.

\section {Do the anomalies depict a coherent picture?}

The important question naturally arises as to whether the anomalies described above
can be simultaneously interpreted within a consistent theoretical framework. 
To this purpose many analyses have been performed in the literature (see for example the
overview given in\cite{Abazajian:2012ys}).  

A first important conclusion derived from these works is that those 
models incorporating more than one sterile neutrino
are disfavored at least for three reasons: I) They are incompatible with primordial nucleosyntesis\cite{Izotov:2010ca,Mangano:2011ar};
II) They introduce an absolute neutrino mass content that cannot be tolerated by 
cosmological data\footnote{This is true even if one takes into account the (small) departures from 
full thermalization\cite{Jacques:2013xr}.}
(indeed, the 3+1 scheme is already borderline in such a respect);
III) Differently from the past, they are no more necessary to explain (through CP violation effects) the 
mismatch between the neutrino and antineutrino excesses previously found in MiniBooNE, and 
now much reduced in the latest data release.

We are thus left with the 3+1 scheme arguably favored by Occam's razor\cite{Giunti:2011gz},
 which however, has its own troubles.
Indeed, interpreted in this framework, the anomalous  reactor and gallium $\nu_e^{\!\!\!\!\!\!(-)}$-disappearance
measurements point towards a non-zero value of the parameter $|U_{e4}|^2$. On the other hand, 
all searches of a possible $\nu_\mu^{\!\!\!\!\!\!(-)}$ disappearance induced by sterile neutrino oscillations
have given negative outcome\cite{Dydak:1983zq,Adamson:2011ku,Cheng:2012yy,Mahn:2011ea}, 
implying a stringent upper bound on the amplitude of $|U_{\mu4}|^2$.
The different results of  the $\nu_e^{\!\!\!\!\!\!(-)}$ and $\nu_\mu^{\!\!\!\!\!\!(-)}$ disappearance searches are perfectly consistent but
(taken together) are in strong tension with the positive signal of $\nu_\mu^{\!\!\!\!\!\!(-)} \to \nu_e^{\!\!\!\!\!\!(-)}$ conversion found at
LSND and MiniBooNE, which requires a (too large) value for the product  $|U_{e4}|^2|U_{\mu4}|^2$
(see the discussion presented in\cite{Giunti:2011gz,Kopp:2011qd}). This situation is depicted 
in Fig.~4, where the constraints from disappearance and appearance data are compared for
the 3+1 model\cite{Giunti:2011cp}. Adding further sterile neutrino species does not help in reducing such a tension, 
which indeed persists also in the more general 3+2 and  3+3 schemes 
(see\cite{Giunti:2011gz,Conrad:2012qt}).

\begin{figure*}[t!]
\vspace*{-0.0cm}
\hspace*{2.00cm}
\includegraphics[width=8.0 cm]{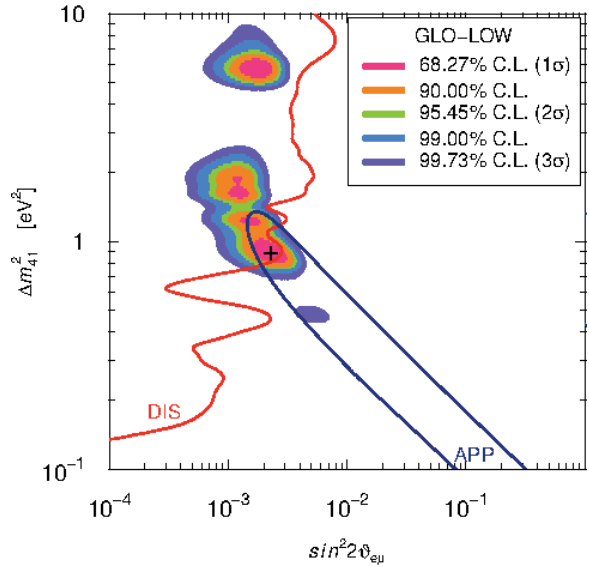}
\vspace*{0.0cm}
\caption{The solid lines delimit the $3\sigma$ regions allowed respectively by the 
appearance and disappearance data, while the colored regions are those preferred
by the combination of the two datasets (figure taken from\protect\cite{Giunti:2011cp}).
\label{fig_tension}}
\end{figure*}  

Concerning the cosmological data, it must be noted that the bounds on the absolute mass
of the new light particles lie in the sub-eV range and therefore they are in tension with the estimates 
coming from the laboratory anomalies,  which point towards a mass-squared splitting of $\sim$\,1 eV$^2$
(see for example\cite{Archidiacono:2012ri}). 

Therefore, the (provisional) answer to the question posed above is negative, as the simultaneous interpretation 
of {\it all} the anomalous data is problematic. However,  we deem it  premature to abandon the sterile neutrino
hypothesis on such a  basis.
Indeed, the possibility exists that some of the experimental results may be wrong while other being correct. 
We remind that the discovery of neutrino oscillations originated in a similarly
confused landscape of discrepant measurements,  which eventually converged in a clear (positive) picture.
 The same thing may happen also in the present case.

\section{An independent test of the indication of $|U_{e4}|^2>0$}

In this perspective it is essential to carefully put under test any single piece of evidence,
irrespective of all the other ones.  Here, we focus on the reactor and gallium anomalies, 
which {\em taken alone} are consistent with a non-zero admixture of the electron neutrino
with new sterile species.

As we have shown in\cite{Palazzo:2011rj},  where we have presented the analytical treatment 
of the solar MSW transitions in a 3+1 scheme (see also\cite{Giunti:2009xz}),
the solar sector data  (Solar and KamLAND) offer a sensitive probe of the admixture 
of the electron neutrino with new sterile species.
In a subsequent paper\cite{Palazzo:2012yf}, we have shown how the first evidence 
for non-zero $\theta_{13}$  further improved
the sensitivity of the solar sector to $U_{e4}$. In this review we present an updated version%
\footnote{Preliminary results of this analysis have been presented in\cite{Palazzo_now2012}.}
of the analysis performed in\cite{Palazzo:2012yf} by incorporating the latest (strongest) constraints 
on $\theta_{13}$, which is now determined with a far better precision. 

In the left panel of Fig.~5, the diagonal bands indicate the region allowed by the combined solar and KamLAND data
in the plane [$\sin^2\theta_{13}, \sin^2\theta_{14}$].%
\footnote{In the parametrization adopted in\cite{Palazzo:2011rj,Palazzo:2012yf} $|U_{e4}|^2 \equiv \sin^2\theta_{14}$.}
We stress that the KamLAND analysis has been performed using only the spectral
shape information so as  to render its results independent of the reactor antineutrino
flux normalization. In the same panel, the vertical bands identify the range allowed for $\theta_{13}$ 
by the combination of the dual-baseline reactor experiments Daya Bay\cite{An:2012bu} and 
RENO\cite{Ahn:2012nd}. In order to understand this behavior, it should be observed that at 
distances of a few hundreds meters, typical of the 
near and far detectors of the dual-baseline reactor experiments,
the oscillations driven by the new mass-squared splitting 
get completely averaged for $\Delta m^2_{14}$ in the region 
of current interest  around $\sim 1~\mathrm{eV}^2$
and their effect (an energy independent suppression) is identical at the near
and far sites. Therefore, the estimate of $\theta_{13}$ (based on a 
near/far comparison) is insensitive to $\theta_{14}$ (as well as to the reactor flux normalization).
The same conclusion is not true for Double Chooz\cite{Abe:2012tg}, currently working only 
with the far detector (see the discussion presented in\cite{Giunti:2011vc}).

\begin{figure*}[t!]
\vspace*{-8.3cm}
\hspace*{-0.4cm}
\includegraphics[width=19.0 cm]{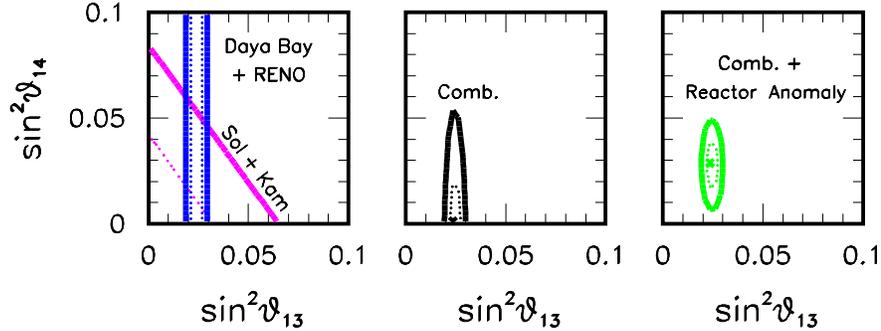}
\vspace*{-6.4cm}
\caption{Left panel: regions allowed by the solar sector data (diagonal bands) and
by the dual-baseline reactor (Daya Bay and RENO) experiments (vertical bands). 
Middle panel: regions allowed by their combination. Right panel: 
combination of the constraints in the middle panel with those coming 
from the reactor anomaly. 
The contours refer to $\Delta \chi^2 =1$ (dotted  line)  and $\Delta \chi^2 = 4$ (solid line).
\label{fig2}}
\end{figure*}  

The superposition of the two datasets (solar+KamLAND and  Daya Bay+RENO)
evidences their complementarity in constraining the two mixing angles.
Their combination, shown in the middle panel of Fig.~5,
leads to the strong upper bound
\begin{equation} 
\sin^2 \theta_{14}\equiv |U_{e4}|^2 < 0.041~~~~ (90\%\, \mathrm {C.L.})\,.
\label{eq:bound}
\end{equation}
As already stressed, this bound does not depend on the normalization of the
reactor fluxes and thus represents an independent constraint
on $|U_{e4}|^2$. Therefore, it makes sense to combine such a bound with
the information coming from the reactor anomaly.%
\footnote{We do not combine the bound in Eq.~(1) with the constraints 
coming from the gallium anomaly 
since they are not independent. Indeed, a shift in the theoretical estimate of the 
cross-section $\nu_e +^{71\!}\!{\mathrm {Ga}} \to ^{71\!}\!{\mathrm {Ge}} + e^-$, which 
is a critical issue for the gallium anomaly, would also modify the solar bound
on $\theta_{14}$ (and $\theta_{13}$).}
The result of such an exercise is shown in the right panel of Fig.~5, where we have taken the
reactor anomaly likelihood from\cite{Giunti:2012tn}.
The effect of the bound in Eq.~(\ref{eq:bound}) is a downshift of the $2\sigma$ range 
allowed for $|U_{e4}|^2$ from $[0.011, 0.054]$\cite{Giunti:2012tn} to $[0.005, 0.050]$, and a sensitive
reduction of the overall statistical significance (from $\sim3\sigma$ to $\sim 2.5\sigma$) 
of the indication of $\theta_{14}>0$.

\section{Conclusions}
 
We have presented a concise discussion of the current phenomenology of light 
sterile neutrinos. The present situation appears quite confused and new experimental
input is indispensable to shed light on the issue. A general consensus towards the absolute necessity 
to perform new and decisive experimental tests has clearly emerged within the ``neutrino
community''. This aspiration is documented by a ``white paper''\cite{Abazajian:2012ys}
coauthored by a large number of scientists (theorists, experimenters and phenomenologists) 
affiliated with more than one hundreds international institutions. 

Such a goal requires the realization of new and more sensitive experiments, able (in case of negative outcome) 
to definitely rule out the sterile neutrino hypothesis, and (in case
of positive outcome) provide a clear observation of the oscillatory pattern 
in the energy and/or space domains, which is the distinctive and indisputable 
signature of the flavor oscillation phenomenon.  

It is important to realize that the new more sensitive experimental tests (see\cite{Abazajian:2012ys}
for a comprehensive overview) will be extremely useful independently of their outcome. 
In case of a positive signal (evidence of sterile neutrino oscillations) we 
would be faced with an extraordinary discovery: new physics beyond the Standard
Model would manifest in a completely unexpected form, well different from the mainstream
high-energy realizations commonly investigated. On the other hand, in case of a negative result, the 
new experiments will be able to put a stringent upper bound on the new 
mass-mixing parameters, thus ruling out definitively the light sterile neutrino hypothesis
as the explanation of the puzzling experimental anomalies.

\section* {Acknowledgments}

We acknowledge support from the European Community through a Marie Curie IntraEuropean 
Fellowship, grant agreement no. PIEF-GA-2011-299582, ``On the Trails of New Neutrino Properties".
We also acknowledge partial support from the  European Union FP7  ITN
INVISIBLES (Marie Curie Actions, PITN-GA-2011-289442).


\begin{thebibliography}{99}


\bibitem{Fogli:2012ua} 
  G.~L.~Fogli, E.~Lisi, A.~Marrone, D.~Montanino, A.~Palazzo and A.~M.~Rotunno,
  Phys.\ Rev.\ D {\bf 86}, 013012 (2012).


\bibitem{ALEPH:2005ab} 
  S.~Schael {\it et al.}  [ALEPH and DELPHI and L3 and OPAL and SLD and LEP Electroweak Working Group and SLD Electroweak Group and SLD Heavy Flavour Group Collaborations],
  Phys.\ Rept.\  {\bf 427}, 257 (2006).


\bibitem{Abazajian:2012ys} 
  K.~N.~Abazajian {\it et al.}, 
  arXiv:1204.5379 [hep-ph].



\bibitem{Athanassopoulos:1996jb} 
  C.~Athanassopoulos {\it et al.}  [LSND Collaboration],
  Phys.\ Rev.\ Lett.\  {\bf 77}, 3082 (1996).
  
\bibitem{Aguilar:2001ty}
  A.~Aguilar {\it et al.}, 
  Phys.\ Rev.\  D {\bf 64}, 112007 (2001).
  
\bibitem{Armbruster:2002mp} 
  B.~Armbruster {\it et al.}  [KARMEN Collaboration],
  Phys.\ Rev.\ D {\bf 65}, 112001 (2002).
 
\bibitem{Church:2002tc} 
  E.~D.~Church, K.~Eitel, G.~B.~Mills and M.~Steidl,
  Phys.\ Rev.\ D {\bf 66}, 013001 (2002).
  
      
\bibitem{AguilarArevalo:2012va} 
  A.~A.~Aguilar-Arevalo {\it et al.},  
  arXiv:1207.4809 [hep-ex].

   

\bibitem{Antonello:2012pq} 
  M.~Antonello {\it et al.},
  arXiv:1209.0122 [hep-ex].



\bibitem{Mueller:2011nm} 
  T.~A.~Mueller {\it et al.},
  Phys.\ Rev.\ C {\bf 83}, 054615 (2011).
  
\bibitem{Huber:2011wv} 
  P.~Huber,
  Phys.\ Rev.\ C {\bf 84}, 024617 (2011)
  [Erratum-ibid.\ C {\bf 85}, 029901 (2012)].



\bibitem{Vogel:1980bk}
  P.~Vogel, G.~K.~Schenter, F.~M.~Mann, R.~E.~Schenter,
  Phys.\ Rev.\  C {\bf 24}, 1543-1553 (1981).

\bibitem{VonFeilitzsch:1982jw}
  F.~Von Feilitzsch, A.~A.~Hahn, K.~Schreckenbach,
  Phys.\ Lett.\  B {\bf 118}, 162-166 (1982).
  
\bibitem{Schreckenbach:1985ep}
  K.~Schreckenbach, G.~Colvin, W.~Gelletly, F.~Von Feilitzsch,
  Phys.\ Lett.\  B {\bf 160}, 325-330 (1985).

\bibitem{Hahn:1989zr}
  A.~A.~Hahn, K.~Schreckenbach, G.~Colvin, B.~Krusche, W.~Gelletly, F.~Von Feilitzsch,
  Phys.\ Lett.\  B {\bf 218}, 365-368 (1989).
  
  
  

\bibitem{Mention:2011rk} 
  G.~Mention {\it et al.}, 
  Phys.\ Rev.\ D {\bf 83}, 073006 (2011).
   
 

 
\bibitem{Abdurashitov:2005tb}
  J.~N.~Abdurashitov {\it et al.},
  Phys.\ Rev.\  C {\bf 73}, 045805 (2006).

\bibitem{Giunti:2012tn} 
  C.~Giunti {\it et al.}, 
  arXiv:1210.5715 [hep-ph].




\bibitem{Komatsu:2010fb} 
  E.~Komatsu {\it et al.}  [WMAP Collaboration],
  Astrophys.\ J.\ Suppl.\  {\bf 192}, 18 (2011).
 
    
\bibitem{Hamann:2010bk}
  J.~Hamann {\it et al.}, 
  Phys.\ Rev.\ Lett.\  {\bf 105}, 181301 (2010).
   
 
\bibitem{Giusarma:2011ex} 
  E.~Giusarma {\it et al.}, 
  Phys.\ Rev.\ D {\bf 83}, 115023 (2011).

\bibitem{Hamann:2011ge} 
  J.~Hamann, S.~Hannestad, G.~G.~Raffelt and Y.~Y.~Y.~Wong,
  JCAP {\bf 1109}, 034 (2011).

\bibitem{Archidiacono:2011gq} 
  M.~Archidiacono, E.~Calabrese and A.~Melchiorri,
  Phys.\ Rev.\ D {\bf 84}, 123008 (2011).
  
\bibitem{Hamann:2011hu} 
  J.~Hamann,
  JCAP {\bf 1203}, 021 (2012).

\bibitem{Joudaki:2012uk} 
  S.~Joudaki, K.~N.~Abazajian and M.~Kaplinghat,
  arXiv:1208.4354 [astro-ph.CO].

\bibitem{Wang:2012vh} 
  X.~Wang, X.~L.~Meng, T.~J.~Zhang, H.~Shan, Y.~Gong, C.~Tao, X.~Chen and Y.~F.~Huang,
  JCAP {\bf 1211}, 018 (2012).
  
\bibitem{Giusarma:2012ph} 
  E.~Giusarma, R.~de Putter and O.~Mena,
  arXiv:1211.2154 [astro-ph.CO].




\bibitem{Izotov:2010ca}
  Y.~I.~Izotov, T.~X.~Thuan,
  Astrophys.\ J.\  {\bf 710}, L67-L71 (2010).

\bibitem{Mangano:2011ar} 
  G.~Mangano and P.~D.~Serpico,
  Phys.\ Lett.\ B {\bf 701}, 296 (2011).


\bibitem{Mirizzi:2012we} 
  A.~Mirizzi, N.~Saviano, G.~Miele and P.~D.~Serpico,
  Phys.\ Rev.\ D {\bf 86}, 053009 (2012).
  
\bibitem{Hannestad:2012ky} 
  S.~Hannestad, I.~Tamborra and T.~Tram,
  JCAP {\bf 1207}, 025 (2012).

\bibitem{Saviano:2013ktj} 
  N.~Saviano, A.~Mirizzi, O.~Pisanti, P.~D.~Serpico, G.~Mangano and G.~Miele,
  arXiv:1302.1200 [astro-ph.CO].
  

\bibitem{Hasenkamp:2012ii} 
  J.~Hasenkamp and J.~Kersten,
  arXiv:1212.4160 [hep-ph].
  


\bibitem{Jacques:2013xr} 
  T.~D.~Jacques, L.~M.~Krauss and C.~Lunardini,
  arXiv:1301.3119 [astro-ph.CO].


\bibitem{Giunti:2011gz} 
  C.~Giunti and M.~Laveder,
  Phys.\ Rev.\ D {\bf 84}, 073008 (2011).




\bibitem{Dydak:1983zq} 
  F.~Dydak {\it et al.}, 
  Phys.\ Lett.\ B {\bf 134}, 281 (1984).
  
\bibitem{Adamson:2011ku} 
  P.~Adamson {\it et al.},  
  Phys.\ Rev.\ Lett.\  {\bf 107}, 011802 (2011).
  
\bibitem{Cheng:2012yy} 
  G.~Cheng {\it et al.},  
  Phys.\ Rev.\ D {\bf 86}, 052009 (2012).
  
\bibitem{Mahn:2011ea}
  K.~B.~M.~Mahn {\it et al.},  
  Phys.\ Rev.\ D {\bf 85} 032007 (2012).
  
  


\bibitem{Kopp:2011qd} 
  J.~Kopp, M.~Maltoni and T.~Schwetz,
  Phys.\ Rev.\ Lett.\  {\bf 107}, 091801 (2011).

\bibitem{Giunti:2011cp} 
  C.~Giunti and M.~Laveder,
  Phys.\ Lett.\ B {\bf 706}, 200 (2011).
  
\bibitem{Conrad:2012qt} 
  J.~M.~Conrad, C.~M.~Ignarra, G.~Karagiorgi, M.~H.~Shaevitz and J.~Spitz,
  arXiv:1207.4765 [hep-ex].
  

\bibitem{Archidiacono:2012ri}
  M.~Archidiacono {\it et al.}, 
  Phys.\ Rev.\ D {\bf 86}, 065028 (2012).
  

\bibitem{Palazzo:2011rj} 
  A.~Palazzo,
  Phys.\ Rev.\ D {\bf 83}, 113013 (2011).

\bibitem{Giunti:2009xz}
  C.~Giunti and Y.~F.~Li,
  Phys.\ Rev.\  D {\bf 80}, 113007 (2009).

\bibitem{Palazzo:2012yf} 
  A.~Palazzo,
  Phys.\ Rev.\ D {\bf 85}, 077301 (2012).

\bibitem{Palazzo_now2012}
A.~Palazzo, Proceedings of NOW 2012,
Neutrino Oscillation Workshop (Conca
Specchiulla, Lecce, Italy, 2012), ed. by P. Bernardini, G.L. Fogli, and E. Lisi, to appear in 
Nucl. Phys. B (Proc. Suppl.)




\bibitem{An:2012bu}
  F.~P.~An {\it et al.},  
  arXiv:1210.6327 [hep-ex].

\bibitem{Ahn:2012nd} 
  J.~K.~Ahn {\it et al.},  
  Phys.\ Rev.\ Lett.\  {\bf 108}, 191802 (2012).


\bibitem{Abe:2012tg} 
  Y.~Abe {\it et al.}, 
  Phys.\ Rev.\ D {\bf 86}, 052008 (2012).

\bibitem{Giunti:2011vc} 
  C.~Giunti and M.~Laveder,
  Phys.\ Rev.\ D {\bf 85}, 031301 (2012).
  

 \end{thebibliography}
\end{document}